\documentclass[twocolumn,aps,prc,superscriptaddress,showpacs,floatfix]{revtex4}
\usepackage{amssymb}
\usepackage{amsmath}
\usepackage{graphicx}
\begin{document}

\title{Production of antimatter $^{5,6}$Li nuclei in central Au+Au collisions at $\sqrt{s_{NN}} = 200$ GeV}

\author{Kai-Jia Sun}
\affiliation{Department of Physics and Astronomy and Shanghai Key Laboratory for
Particle Physics and Cosmology, Shanghai Jiao Tong University, Shanghai 200240, China}
\author{Lie-Wen Chen\footnote{%
Corresponding author (email: lwchen$@$sjtu.edu.cn)}}
\affiliation{Department of Physics and Astronomy and Shanghai Key Laboratory for
Particle Physics and Cosmology, Shanghai Jiao Tong University, Shanghai 200240, China}
\affiliation{Center of Theoretical Nuclear Physics, National Laboratory of Heavy Ion
Accelerator, Lanzhou 730000, China}
\date{\today}

\begin{abstract}
Combining the covariant coalescence model and a blast-wave-like analytical
parametrization for (anti-)nucleon phase-space freezeout configuration,
we explore light (anti-)nucleus production in central
Au+Au collisions at $\sqrt{s_{NN}} = 200$ GeV.
Using the nucleon freezeout configuration (denoted by FO1) determined from
the measured spectra of protons (p), deutrons (d) and $^{3}$He, we find the
predicted yield of $^{4}$He is significantly smaller than the experimental
data. We show this disagreement can be removed by using a nucleon freezeout
configuration (denoted by FO2) in which the nucleons are assumed to freeze
out earlier than those in FO1 to effectively consider the effect of large
binding energy value of $^{4}$He.
Assuming the binding energy effect also exists for the production of
$^5\text{Li}$, $^5\overline{\text{Li}}$, $^6\text{Li}$ and $^6\overline{\text{Li}}$
due to their similar binding energy values as $^{4}$He, we find the
yields of these heavier (anti-)nuclei can be enhanced by a factor
of about one order, implying that although the stable (anti-)$^6$Li
nucleus is unlikely to be observed, the unstable (anti-)$^5$Li nucleus
could be produced in observable abundance in Au+Au collisions at
$\sqrt{s_{NN}} = 200$ GeV where it may be identified through the p-$^4\text{He}$
($\overline{\text{p}}$-$^4\overline{\text{He}}$) invariant mass spectrum.
The future experimental measurement on (anti-)\text{$^5\text{Li}$} would
be very useful to understand the production mechanism of heavier antimatter.
\end{abstract}

\pacs{25.75.-q, 25.75.Dw}
\maketitle

\section{Introduction}

\label{introduction}

The quest for antimatter has become one of fundamental issues in contemporary
physics, astronomy and cosmology since the discovery of the positron (the antielectron)
in cosmic radiation~\cite{And33} which corresponds to the negative energy states
of electrons predicted by Dirac~\cite{Dir28}. Based on very general principles
of relativistic quantum field theory, it is believed that each particle has its
corresponding antiparticle of the same mass (but the opposite charge) and any physical
system has an antimatter analog with an identical mass. Indeed, following the
observation of antiprotons ($\overline{\text{p}}$)~\cite{Cha55} and antineutrons ($\overline{\text{n}}$)~\cite{Cor56},
more complex antimatter nuclei such as antideutrons ($\overline{\text{d}}$)~\cite{Mas65,Dor65},
antihelium-3 ($^3\overline{\text{He}}$)~\cite{Ant70} and antitritons ($^3\overline{\text{H}}$)~\cite{Vis74}
have been observed. In terrestrial laboratories,
the antihydrogen atoms have also been produced~\cite{Bau96} and can even survive
for a long time in confinement~\cite{ALPHA11}.
Recently, STAR collaboration
at RHIC reported the discovery of strange antimatter nucleus,
the antihypertriton ($^3_{\overline{\Lambda}}\overline{\text{H}}$)~\cite{Abe10},
and the heavier antimatter nucleus antihelium-4
($^4\overline{\text{He}}$ or $\overline{\alpha}$)~\cite{Abe11} in Au+Au collisions.
The ALICE collaboration at LHC also claimed the observation of $^4\overline{\text{He}}$ in Pb+Pb
collisions~\cite{Sha11}. A recent review on antimatter production can be found in Ref.~\cite{Ma12}.

The study of antimatter nuclei production in heavy-ion collisions is of critical
importance for a number of fundamental problems in physics, astronomy and cosmology.
For example, the precision measurement of the mass difference between nuclei and
anti-nuclei can test the fundamental CPT theorem for systems bound by the strong
interaction~\cite{Ada15}. The measured production rate of light anti-nuclei in heavy-ion collisions
provides a point of reference for possible future observations in cosmic radiation
for the motivation of hunting for antimatter and dark matter in the Universe~\cite{Ahl94,Fuk05,Don08,Abe12}.
The antimatter nuclei production provides the possibility to test the interactions
between antimatter and antimatter~\cite{Ada15}.
In addition, the production of light anti-nuclei in heavy-ion collisions can be used
to extract the freezeout information of antinucleons in these collisions, which is useful
to infer the properties of a new state of matter, i.e., quark-gluon plasma (QGP)
possibly formed in these collisions
as well as to understand how the QGP expands, cools and
hadronizes, providing a new window (compared to the electromagnetic and
hadronic probes) for exploring the dynamics of ultrarelativistic heavy-ion collisions.

The heaviest antimatter nucleus observed so far is $^4\overline{\text{He}}$, and it will remain the
heaviest stable antimatter nucleus observed for the foreseeable future~\cite{Abe11} barring some dramatic
discoveries in space detectors due to some special production mechanism~\cite{Gre96} or a new breakthrough in
accelerator technology. This is because the (anti-)nucleus production rate in these heavy-ion collisions
is found to reduce by a factor of about $10^3$ for each additional (anti-)nucleon added to the (anti-)nucleus
according to the measured yields of $\overline{\text{p}}$ (p), $\overline{\text{d}}$ (d),
$^3\overline{\text{He}}$ ($^3$He) and $^4\overline{\text{He}}$ ($^4$He)~\cite{Abe11},
and thus the yield of the next heavier stable antimatter nucleus, antilithium-6 ($^6\overline{\text{Li}}$), is
expected to be down by a factor of about $10^6$ compared to $^4\overline{\text{He}}$ assuming
$^6\overline{\text{Li}}$ production rate follows the same exponential reduction law and
is beyond the reach of current accelerator technology. However, a careful observation on the
experimental yields of p ($\overline{\text{p}}$), d ($\overline{\text{d}}$),
$^3$He ($^3\overline{\text{He}}$) and $^4$He ($^4\overline{\text{He}}$) in central Au+Au
collisions at $\sqrt{s_{NN}} = 200$ GeV from STAR~\cite{Abe11} indicates that while the yields of
p ($\overline{\text{p}}$), d ($\overline{\text{d}}$) and $^3$He ($^3\overline{\text{He}}$)
follow an exponential reduction rate very well, the yield of $^4$He ($^4\overline{\text{He}}$)
displays a significant enhancement (excess) compared to the exponential reduction rate.
The coalescence model calculations also significantly underestimate the yield of
$^4$He ($^4\overline{\text{He}}$) although they can successfully describe the yields of
p ($\overline{\text{p}}$), d ($\overline{\text{d}}$) and $^3$He ($^3\overline{\text{He}}$)~\cite{Xue12}.
It is thus of great
interest to understand the physics behind this enhancement for the yield of
$^4$He ($^4\overline{\text{He}}$), which would be critically important for the
future searching for heavier antimatter nuclei such as antilithium-5 ($^5\overline{\text{Li}}$)
and $^6\overline{\text{Li}}$ in heavy-ion collisions.
In Ref.~\cite{Gre96}, the possibility of direct production of antimatter
nuclei out of the highly correlated vacuum has been discussed, which provides
a potentially more copious production mechanism for heavier antimatter nuclei
in heavy-ion collisions.

In the present work, we propose that
the enhancement of the $^4$He ($^4\overline{\text{He}}$) yield could be
due to its large binding energy which leads to relatively earlier formation
for $^4$He ($^4\overline{\text{He}}$) than for d ($\overline{\text{d}}$) and
$^3$He ($^3\overline{\text{He}}$) in the heavy-ion collisions.
Assuming the similar binding energy effects also exist for the production of
$^5$Li, $^5\overline{\text{Li}}$, $^6$Li and $^6\overline{\text{Li}}$ , we find the
predicted yields of these heavier (anti-)nuclei can be enhanced significantly,
implying that
although the stable (anti-)$^6$Li nucleus is unlikely to be observed, the unstable
(anti-)$^5$Li nucleus could be produced in observable abundance in Au+Au collisions
at RHIC.

\section{The theoretical models}
\label{formulism}

Understanding particle production in heavy-ion collision at different energy
regions is among the fundamental questions in nuclear and particle physics.
Theoretically, the microscopic
coalescence model~\cite{But61,Sat81,Cse86}
and the macroscopic thermal model~\cite{Cle91,Bec97,Bra07,And11,Cle11} provide two important
approaches to describe the light cluster production in heavy-ion collisions.
In particular, these two approaches have been successfully applied recently to
describe the production of light (anti-)nuclei in ultrarelativistic heavy-ion collisions at RHIC and LHC energies~\cite{Zha10,Xue12,ChenG12,ChenG13,And11,Cle11,Pop10,Ste12,Cha14,Cha15}.
In the present work,
the theoretical formulism for the description of (anti-)nuclei
production in heavy-ion collisions is based on the covariant coalescence model~\cite{Dov91}
together with the (anti-)nucleon phase-space freezeout configuration described by a
blast-wave-like analytical parametrization~\cite{Ret04} 
which has been shown to be very successful to describe the hadron phase-space
freezeout configuration in ultrarelativistic heavy-ion collisions.

\subsection{(Anti-)Nucleon phase-space freezeout configuration}

One basic ingredient of the coalescence model is the emission source function, i.e.,
the phase-space freezeout configuration, of the constituent particles. In principle,
the phase-space freezeout configuration can be obtained dynamically from transport
model simulations for heavy-ion collisions (see, e.g., Refs.~\cite{Mat97,ChenLW03,ChenLW06,Oh09}). In the present
work, for simplicity, we describe the (anti-)nucleon phase-space freezeout configuration
using a fireball-like model through a blast-wave-like analytical
parametrization~\cite{Ret04}.

We assume that particles are emitted from a freezeout hypersurface $\Sigma^\mu$
where the particles are in local thermal equilibrium described by Lorentz invariant
one-particle distribution function $f(x,p)$ given by~\cite{Coo74}
\begin{eqnarray}
f(x,p)&=&gh^{-3}[\exp((p^{\mu} u_{\mu}-\Omega)/kT)\pm 1]^{-1}   \notag \\
&=&g(2\pi)^{-3}[\exp(p^{\mu}u_{\mu}/kT)/\xi \pm 1]^{-1},
\end{eqnarray}
where the reduced Planck constant $\hbar=\frac{h}{2\pi}$ is set to be $1$, $g$ is spin degeneracy factor,
$\Omega$ is the chemical potential, $\xi=exp(\Omega /kT)$ is the fugacity
which is directly related to particle number density, $u_{\mu}$ is the four-velocity of a fluid element
in the fireball and $T$ is the corresponding local temperature.
For the phase-space freezeout configuration,
instead of using the four coordinates $(t,x,y,z)$, it is convenient to
use the cylindrical coordinates $(\tau,r,\phi_s,\eta)$ (see, e.g., Ref.~\cite{Sch99})
where $\tau =\sqrt{t^{2}-z^{2}}$ is the longitudinal proper time,
$\eta =\frac{1}{2}\ln (\frac{t+z}{t-z})$ is the longitudinal space-time rapidity,
$r$ is the transverse radius, and $\phi_s$ is the spatial azimuthal angle.
Similarly, the four momentum $(E,p_x,p_y,p_z)$ is transformed to
$(m_T,p_T,\phi_p,y)$ where $p_T=\sqrt{p_x^2+p_y^2}$ is the transverse momentum,
$y=\frac{1}{2}\ln (\frac{E+p_z}{E-p_z})$ is rapidity,
$m_T=\sqrt{m^2+p_T^2}$ is transverse mass, and $\phi_p$ is the azimuthal angle in momentum space.
The four coordinate and momentum can thus be expressed as
$x^{\mu}=(\tau \cosh \eta,r\cos\phi _s,r\sin\phi _s,\tau \sinh \eta)$ and
$p^\mu=(m_T\cosh y,p_T\cos\phi_p,p_T\sin\phi_p,m_T\sinh y)$, respectively.

The one-particle invariant momentum distribution can be obtained as
\begin{equation}\label{4}
    E\frac{\text{d}^3N}{\text{d}p^3}=\frac{\text{d}^3N}{p_T\text{d}p_T\text{d}y\text{d}\phi_p}=\int\limits_{\Sigma^\mu}\text{d}^3
    \sigma _{\mu} p^\mu f(x,p),
\end{equation}
with $\text{d}p^3=Ep_T\text{d}p_T\text{d}y \text{d}{\phi_p}$.
For the particle production at midrapidity in heavy-ion collisions that
we are considering in this work, we adopt the longitudinal
boost invariance assumption~\cite{Bjo83}. By setting the longitudinal flow
velocity $\nu_L = z/t$, the longitudinal flow rapidity
$\eta_{flow}=\frac{1}{2}\ln[(1+\nu_{L})/(1-\nu_{L})]$ will be identical to
the space-time rapidity $\eta=\frac{1}{2}\ln[(t+z)/(t-z)]$, and thus the four-velocity
can be expressed as
\begin{eqnarray}
  u^\mu=&&\cosh\rho(r,\phi_s)(\text{cosh}\eta,\tanh\rho (r,\phi_s)\cos\phi_b, \notag \\
 &&\tanh\rho(r,\phi_s)\sin\phi_b,\sinh\eta),
\end{eqnarray}
where $\rho$ is the transverse rapidity of a fluid element in the fireball. The above
expression can be obtained by a longitudinal boost with velocity $\tanh \eta $
multiplied by a transverse boost with velocity $\tanh \rho$~\cite{Ret04}.
If we fix the freezeout hypersurface $\Sigma^\mu$ by choosing a constant proper
time $\tau $, i.e., $\Sigma^\mu$ is independent of the transverse
coordinates, then the covariant normal vector can be expressed as
\begin{eqnarray}
  \text{d}^3\sigma_{\mu}=(\cosh({\eta}),0,0,-\sinh({\eta})){\tau}r\text{d}r\text{d}{\eta}\text{d}{\phi}_s.
\end{eqnarray}
Therefore, one can obtain
\begin{eqnarray}
    p^{\mu} u_{\mu}&=&m_T \cosh\rho \cosh(\eta -y)-p_T \sinh\rho \cos(\phi_p -\phi_b), \notag \\
    p^{\mu} \text{d}^3\sigma_{\mu}&=&\tau m_T \cosh(\eta -y)\text{d}\eta r\text{d}r\text{d}\phi_s,
\end{eqnarray}
where $\phi_b$ is azimuthal direction of the transverse flow~\cite{Ret04}.
Considering that the freezeout
can happen in some time interval, a Gaussian distribution for the freezeout
proper time is introduced as follows
\begin{eqnarray}
J(\tau)=\frac{1}{\Delta \tau \sqrt{2\pi}}\exp(-\frac{(\tau-\tau_0)^2}{2(\Delta \tau)^2}),
\end{eqnarray}
which satisfies
\begin{eqnarray*}
\int J(\tau) \text{d}\tau=1,~~
\int \tau J(\tau) \text{d}\tau=\tau_0,
\end{eqnarray*}
where $\tau_0$ is the mean value of $\tau$ and $\Delta \tau$ is the dispersion
of the $\tau$ distribution function. Therefore, the momentum distribution can
be obtained as
\begin{eqnarray}\label{12}
\frac{\text{d}^3N}{p_T\text{d}p_T\text{d}y\text{d}\phi_p}=&&\int\limits_{\Sigma^\mu} m_T\cosh(\eta-y)
  f(x,p)J(\tau) \notag \\
  &&\tau \text{d}\tau\text{d}\eta r \text{d}r \text{d}\phi_s.
\end{eqnarray}

Following Ref.~\cite{Ret04}, we parameterize the transverse rapidity of a fluid element
in the fireball as
 \begin{eqnarray}
    \rho=\rho_0 \tilde{r}[1+\epsilon \cos(2\phi_b)],
 \end{eqnarray}
where $\rho_0$ is the isotropic part of the transverse flow, $\epsilon$ is the anisotropic part, $\phi_b$
is azimuthal direction of the transverse flow which is not identical to spatial azimuthal angle $\phi_s$, and
$\tilde{r}$ is the ``normalized elliptical radius"
\begin{eqnarray}
    \tilde{r} = \sqrt{\frac{[r\cos{\phi_s}]^2}{R_x^2}+\frac{[r\sin{\phi_s}]^2}{R_y^2}},
\end{eqnarray}
with
\begin{eqnarray}
    \tan{\phi_s} = (\frac{R_y}{R_x})^2\tan{\phi_b},
\end{eqnarray}
where $R_x=R_0(1+s_2)$ is the minor axis of the ellipse, $R_y=R_0(1-s_2)$ is the major axis, and $s_2$ is the geometric anisotropy.
Therefore, the transverse rapidity can also be written as
\begin{equation}\label{param}
   \rho=\rho_0\sqrt{\frac{[r\cos \phi_s]^2}{[R_x^2]}+\frac{[rsin\phi_s]^2}{R_y^2}}[1+\epsilon\cos{2\phi_b}].
\end{equation}

For midrapidity region ($y=0$) in central heavy-ion collisions that
we are considering here, one has $s_2 = \epsilon =0$ and $\phi_b=\phi_s$,
and thus the invariant distribution function can be expressed as
\begin{eqnarray}\label{17}
f(x,p)=&&\frac{g}{(2\pi)^3}\bigg[\exp\big((m_T \cosh\rho \cosh(\eta)- \notag \\
                         &&p_T \sinh\rho \cos(\phi_p -\phi_s))/kT\big)/\xi \pm 1\bigg]^{-1},
\end{eqnarray}
with $\rho=\rho_0 r/R_0$.
One can thus use formula (\ref{12}) and (\ref{17}) to calculate
transverse momentum distribution of midrapidity particles
in central heavy-ion collision.

\subsection{Covariant coalescence model}

In this work, we calculate light (anti-)nucleus production in
ultrarelativistic heavy-ion collisions using the covariant coalescence
formulism~\cite{Dov91}. In the coalescence model, the probability
for producing a nucleus is determined by the overlap of its Wigner
phase-space density with the nucleon phase-space distribution at
freezeout. We consider that $M$ nucleons are combined to form one
nucleus and the total multiplicity of the nucleus can be obtained as
\begin{eqnarray} \label{18}
N_c&=&g_c\int (\prod_{i=1}^{M}\text{d}N_i)\rho_c^W(x_1,...,x_M;p_1,...,p_M) \notag \\
&=&g_c\int \bigg(\prod_{i=1}^{M}d\tau_i J(\tau_i)   p_{i}^\mu\text{d}^3\sigma_{i\mu}\frac{\text{d}^3p_i}{E_i}f(x_i,p_i)\bigg)\times \notag \\
   &&\rho_c^W(x_1,...,x_M;p_1,...,p_M),
\end{eqnarray}
where $\rho_c^W(x_1,...,x_M;p_1,...,p_M)$ is the Wigner density function which gives the coalescence probability,
$g_c$ is the coalescence factor~\cite{Sat81}. By inserting $\delta$ function to conserve momentum, the invariant
differential transverse momentum distribution of the nucleus becomes
\begin{eqnarray}
   E\frac{d^3N_c}{d^3P}&=&Eg_c\int  \bigg(\prod_{i=1}^{M}d\tau_i J(\tau_i) p_{i}^\mu\text{d}^3\sigma_{i\mu}\frac{\text{d}^3p_i}{E_i}f(x_i,p_i)\bigg)\times \notag \\
   &&\rho_c^W(x_1,...,x_M;p_1,...,p_M)\delta^3(\mathbf{P}-\sum_{i=1}^M\mathbf{p_i}).
\end{eqnarray}
The above formula is Lorentz invariant~\cite{Dov91} and the Wigner function is a Lorentz scalar.

For the Wigner function $\rho_c^W(x_1,...,x_M; p_1,...,p_M)$,
following Ref.~\cite{Mat97,ChenLW06}, instead of calculating
it directly using four dimensional coordinators $x_i$ and four dimensional momenta
$p_i$ of the constituent nucleons, we calculate it in the rest frame of
the nucleus. To do so, a Lorentz transformation is performed to obtain the
space-time and energy-momentum coordinates of each nucleon in the rest frame of
the nucleus. To determine the spatial coordinates of the nucleons at equal time
in the rest frame of the nucleus, i.e.,  $\mathbf{r}_{1}, \mathbf{r}_{2}, ..., \mathbf{r}_{M}$,
the nucleons that freeze out earlier are allowed to propagate
freely with constant velocity given by the ratio of their momentum and energies
in the rest frame of the nucleus, until the time when the last nucleons in the
nucleus freezes out. Furthermore, in order to calculate the Wigner function, Jacobi
coordinate is adopted by a transformation of the coordinate as follows~\cite{Mat97,ChenLW03,ChenLW06,ChenLW0407}
\begin{eqnarray}
  \left(
  \begin{array}{c}
\mathbf{R} \\
\mathbf{q}_{1} \\
\cdot \\
\cdot \\
\cdot \\
\mathbf{q}_{M-1}%
\end{array} \right )
=J_{M}\left(
\begin{array}{c}
\mathbf{r}_{1} \\
\mathbf{r}_{2} \\
\cdot \\
\cdot \\
\cdot \\
\mathbf{r}_{M}%
\end{array}%
\right),
\end{eqnarray}
where $\ \ \mathbf{R=}\frac{\sum_{j=1}^{M}m_{j}\mathbf{r}_{j}}{%
\sum_{j=1}^{M}m_{j}}$ is the center-of-mass position vector of the nucleus
and
$\mathbf{q}_{i}=\sqrt{\frac{i}{i+1}}(\frac{\sum_{j=1}^{i}m_{j}\mathbf{r}_{j}}{\sum_{j=1}^{i}m_{j}}-\mathbf{r}_{i+1})$
is the relative coordinate vector.
Correspondingly, in the momentum space, one has
\begin{eqnarray}
  \left(
  \begin{array}{c}
\mathbf{P} \\
\mathbf{k}_{1} \\
\cdot \\
\cdot \\
\cdot \\
\mathbf{k}_{M-1}%
\end{array} \right )
=(J_{M}^{-1})^T\left(
\begin{array}{c}
\mathbf{p}_{1} \\
\mathbf{p}_{2} \\
\cdot \\
\cdot \\
\cdot \\
\mathbf{p}_{M}%
\end{array}%
\right),
\end{eqnarray}
where $\mathbf{P}$ is the total momentum of the nucleus and
$\mathbf{k}_i$ is the relative momentum vector.
The determinant of the Jacobi matrix
is $|J_{M}|=1/\sqrt{M}$, and one then has the following identity
\begin{eqnarray}
  \prod_{i=1}^M\text{d}^3x_i\text{d}^3p_i=\text{d}^3 R\text{d}^3P\prod_{i=1}^{M-1}\text{d}^3q_i\text{d}^3k_i.
\end{eqnarray}

Furthermore, we assume the harmonic wave function for all the light (anti-)nuclei in
the rest frame except the (anti-)deutrons for which we use the well-known Hulth\'{e}n
wave function (see, e.g., Refs.~\cite{Mat97,ChenLW03}). The Wigner function of the
nucleus can then be obtained as~\cite{ChenLW06}
\begin{eqnarray}
&&\rho_c^W(x_1,...,x_M;p_1,...,p_M)                             \notag \\
&=& \rho ^{W}(q_{1},\cdot \cdot \cdot ,q_{M-1},k_1,\cdot
\cdot \cdot ,k_{M-1}) \notag \\
&=& 8^{M-1}\exp \big[-\sum_{i=1}^{M-1}(q_{i}^{2}/\sigma _{i}^{2}+\sigma
_{i}^{2}k_i^{2})\big],
\end{eqnarray}
with $\sigma_i^2 = (m_i w)^{-1}$ where the harmonic oscillator frequency
$\omega$ is related to the root-mean-square (rms) radius of the nucleus
as follows
\begin{eqnarray}
\left \langle r_{M}^{2}\right\rangle =\frac{3}{2M}\frac{1/\omega}{%
  \underset{i=1}{\overset{M}{\sum }}m_{i}   }\underset{i=1}{%
\overset{M}{\sum }}\left[ m_{i}  \left( \underset{j=i+1}{\overset{M}{%
\sum }}\frac{1}{m_{j}}+\underset{j=1}{\overset{i-1}{\sum }}\frac{1}{m_{j}}%
\right) \right].
\end{eqnarray}
Therefore, $\sigma_i ^2$ can be determined by $\left \langle r_M^2 \right\rangle$.
In the case of $m_{1}=m_{2}=\cdot \cdot \cdot \cdot \cdot \cdot =m_{M}=m$,
one can obtain $\sigma^2 = \frac{2M}{3(M-1)}\left \langle r_M^2 \right\rangle$.

\section{result and discussion}

\subsection{(Anti-)Nucleon freezeout configuration from light (anti-)nuclei production}

We focus on the midrapidity light (anti-)nuclei production in central
Au+Au collisions at $\sqrt{s_{NN}}=200$ GeV in this work.
In this case, there are totally six parameters in the blast-wave-like analytical
parametrization for (anti-)nucleon phase-space freezeout configuration, namely,
the kinetic freeze-out temperature $T$, the transverse rapidity $\rho_0$, the
longitudinal mean proper time $\tau_0$, the time dispersion $\Delta \tau$, the transverse
size at freeze-out $R_0$, and the fugacity of particle $\xi$.

For proton phase-space freezeout configuration, we obtain the local temperature
$T = 111.6$ MeV, the transverse rapidity $\rho_0 = 0.978$, and a constraint on
the combination of the proton fugacity $\xi_p$, $\tau_0$, $\Delta \tau$ and $R_0$,
by fitting the measured
spectrum of protons in Au+Au collisions at $\sqrt{s_{NN}}=200$ GeV for $0$-$5\%$
centrality~\cite{Adl04}. To extract the values of $\xi_p$, $\tau_0$, $\Delta \tau$ and $R_0$,
we further fit the measured spectra of deuterons and $^3$He~\cite{Abe09} simultaneously using the results from
the coalescence model (see the Subsection~\ref{ResultsNuclei} for the details),
which leads to $R_0=15.6$ fm, $\tau_0=10.55$ fm/c, $\Delta \tau=3.5$ fm/c and $\xi_p=10.45$.
For antiprotons, we assume they have the same phase-space freezeout configuration as
protons except the fugacity is reduced to $\xi_{\overline p}=7.84$ to describe the measured
yield ratio $\bar{p}/p=0.75$~\cite{Adl04}.
Table~\ref{Parameters} summarizes the parameters of the blast-wave-like
analytical parametrization for (anti-)nucleon phase-space freezeout configuration (denoted as FO1).
It should be pointed out that
we have neglected the difference between protons and neutrons
(antiprotons and antineutrons) for the phase-space freezeout configuration
due to the small isospin chemical potential at freezeout in Au+Au collisions
at $\sqrt{s_{NN}}=200$ GeV~\cite{And11}.
Based on the freezeout configuration of (anti-)nucleons, one can then
predict the production of light (anti-)nuclei using the coalescence model.

\begin{table}
\caption{Parameters of the blast-wave-like analytical parametrization
for (anti-)nucleon phase-space freezeout configuration.}
\begin{tabular}{c|ccccccc}
        \hline \hline
          & T(MeV) & $\rho_0$ & $R_0$ (fm) & $\tau_0$ (fm/c)& $\Delta \tau$ (fm/c)& $\xi_p$ & $\xi_{\overline p}$  \\
         \hline
        FO1 & 111.6  & 0.98 & 15.6 & 10.55 & 3.5 & 10.45 & 7.84  \\
        FO2 & 111.6  & 0.98 & 12.3 & 8.3 & 3.5 & 21.4 & 16.04  \\
        \hline  \hline

\end{tabular}
\label{Parameters}
\end{table}

\subsection{The production of light (anti-)nuclei}
\label{ResultsNuclei}

We use the coalescence model described above to calculate the production of light (anti-)nuclei.
In the coalescence model, the statistical factor $g_c$ is quite important and it is given
by $g_c = \frac{2j+1}{2^N}$~\cite{Sat81} with $j$ and $N$ being, respectively, the spin and
the nucleon number of the nucleus. The spins of d, $^3$He, $^4$He, $^5$Li and $^6$Li
are $1$, $1/2$, $0$, $3/2$ and $1$, respectively. Furthermore, the rms
radius $r_{\text{rms}}$ of the light nucleus is also important since it determines the
harmonic oscillator frequency parameter $\omega $ in the Wigner function of the nucleus.
The $r_{\text{rms}}$ of d, $^3$He, $^4$He, $^5$Li and $^6$Li are taken to be $1.96$ fm,
$1.76$ fm, $1.45$ fm, $2.5$ fm and $2.5$ fm, respectively~\cite{Rop09,Tan85}.
Here the $r_{\text{rms}} = 2.5$ fm for $^5$Li is estimated based
on the work in Ref.~\cite{Tan85}. For the antinuclei, we assume they have the same ground
state properties as their corresponding nuclei.
Table~\ref{C_factor} summarizes the statistical factors, rms radii as well as the binding
energies~\cite{Wan12} of different light (anti-)nuclei.
It should be mentioned that
while d ($\overline{\text{d}}$), $^3$He ($^3\overline{\text{He}}$),
$^4$He ($^4\overline{\text{He}}$) and $^6$Li ($^6\overline{\text{Li}}$) are stable,
$^5$Li ($^5\overline{\text{Li}}$) is unstable against the proton (antiproton) decay
with half-life of about $370\times 10^{-24}$ s (i.e., $111$ fm/c)~\cite{Aud12} and
thus it may be identified through the p-$^4\text{He}$
($\overline{\text{p}}$-$^4\overline{\text{He}}$) invariant mass spectrum
in heavy-ion collisions.

\begin{table}
\caption{Statistical factor $g_c$, root-mean-square radii $r_{\text{rms}}$~\cite{Rop09,Tan85} and
binding energy $E_b$~\cite{Wan12} of light (anti-)nuclei}

\begin{tabular}{c|c|c|c|c|c}
        \hline \hline
            & d ($\overline{\text{d}}$) & $^3$He ($^3\overline{\text{He}}$) & $^4$He ($^4\overline{\text{He}}$) & $^5$Li ($^5\overline{\text{Li}}$) & $^6$Li ($^6\overline{\text{Li}}$)  \\
         \hline
        $g_c$ & $\frac{(2\times1+1)}{2^2}$ & $\frac{(2\times{\frac{1}{2}}+1)}{2^3}$ &
        $\frac{(2\times{0}+1)}{2^4}$ & $\frac{(2\times{\frac{3}{2}}+1)}{2^5}$ & $\frac{(2\times{1}+1)}{2^6}$  \\
        \hline
         $r_{\text{rms}}$ (fm)   & 1.96  & 1.76  & 1.45  & 2.5 & 2.5  \\
        \hline
         $E_b$ (MeV)   & 2.224  & 7.718  & 28.296  & 26.330 & 31.994  \\
        \hline \hline
\end{tabular}
\label{C_factor}
\end{table}

\begin{figure}
\includegraphics[scale=0.4]{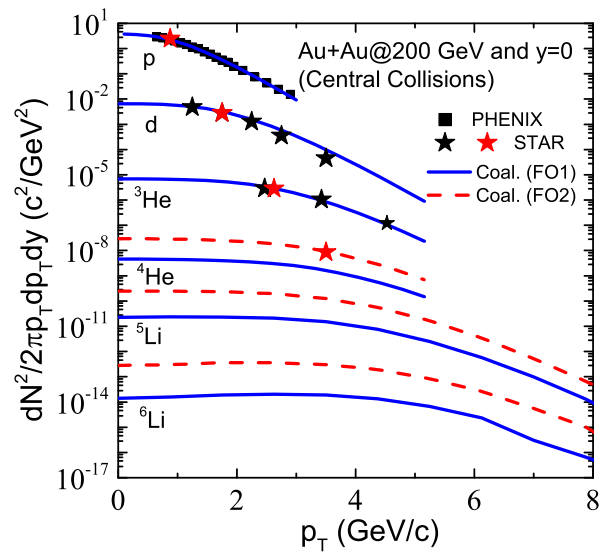}
\caption{Transverse momentum distributions of
light nuclei at midrapidity (y=0) in central Au+Au collisions at
$\protect\sqrt{s_{NN}}=200$ GeV predicted by coalescence model
with FO1 (solid lines) and FO2 (dashed lines). The experiment data
of protons is taken from the PHENIX
measurement~\cite{Adl04} whereas those of light nuclei
are from the STAR measurement~\cite{Abe11,Abe09}.
The data point of protons from STAR measurement has been scaled by a
factor of $0.6$ to correct the weak decay effects~\cite{Ada04}.}
\label{Spect}
\end{figure}

Figure~\ref{Spect} shows the predicted midrapidity transverse momentum
distributions of p, d, $^3\text{He}$, $^4\text{He}$, $^5\text{Li}$ and
$^6\text{Li}$ together with the experimental data of p from PHENIX
collaboration~\cite{Adl04} and the data of p, d, $^3\text{He}$ and
$^4\text{He}$ from STAR collaboration~\cite{Abe09,Abe11,Ada04}
in central Au+Au collisions at $\sqrt{s_{NN}}=200$ GeV.
It is seen that the coalescence model predictions with the freezeout
configuration FO1 are in very good agreement with the measured
transverse momentum distributions of p, d and $^3\text{He}$ as
expected but significantly underestimate the measured yield of
$^4\text{He}$ by a factor of about $6$. The similar feature was
also observed in the calculations in Ref.~\cite{Xue12}.

From Table~\ref{C_factor}, one can see that $^4\text{He}$ has a
specially larger binding energy value compared to d or $^3\text{He}$,
and thus it is more tightly bound and could be formed in relatively
earlier stage in heavy-ion collisions compared to d or $^3\text{He}$.
Physically, the light nuclei
can be formed in principle in the whole dynamical process of heavy-ion
collisions, but they are usually destroyed immediately after their formation
due to the violating collisions in the high temperature environment.
However, for the light nuclei with large binding energy values such as
$^4\text{He}$, the survival probability in relatively
earlier stage in heavy-ion collisions is expected to enhance
compared to the loosely bound d and $^3\text{He}$.
In principle, these effects can be studied using
transport model simulations with dynamic light cluster production
in heavy-ion collisions~\cite{Dan91} although this
is highly nontrivial and beyond the scope of this work.
In the present work, to effectively mimic
this binding energy effect, we assume the volume (time) of the freezeout
hypersurface for nucleons coalesced into $^4\text{He}$ is smaller (shorter)
than that of d or $^3\text{He}$.
For simplicity, we reduce $R$ and $\tau$ by a factor $1.27$ to
fit the measured yield of $^4\text{He}$, and this leads
to $R=12.3 $ fm, $\tau=8.3$ fm/c, $\Delta \tau=3.5$ fm/c and
$\xi_p=21.4$, which is denoted as the phase-space freezeout
configuration FO2 and is summarized in Table~\ref{Parameters}.
Since the binding energy values
of $^5\text{Li}$ and $^6\text{Li}$ are also large and comparable
with that of $^4\text{He}$ as shown in Table~\ref{C_factor},
it is thus expected that the nucleons coalesced into $^5\text{Li}$ and
$^6\text{Li}$ should have similar phase-space freezeout configuration as those
coalesced into $^4\text{He}$.
In Fig.~\ref{Spect}, we also include the predicted transverse
momentum distributions of $^4\text{He}$, $^5\text{Li}$ and
$^6\text{Li}$ with the freezeout configuration FO2.
It is seen that using the freezeout configuration FO2
significantly enhance the yields of $^5\text{Li}$ and
$^6\text{Li}$ by a factor of about $9$ and
$16$, respectively, compared to the case using FO1.

\begin{figure}
\includegraphics[scale=0.43]{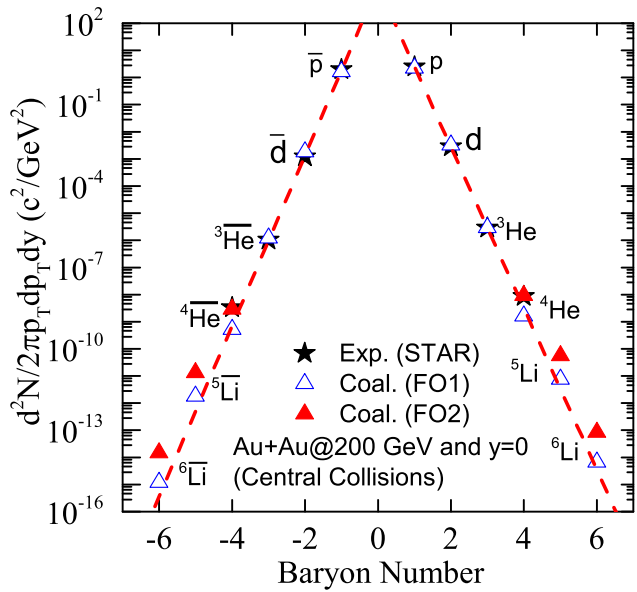}
\caption{The differential invariant yields $d^2N/(2\pi p_Tdp_Tdy)$ of
(anti-)nucleus at the transverse momentum $p_T/|B|=0.875$ GeV/c as a
function of baryon number $B$ in central Au+Au collisions
at $\sqrt{s_{NN}}=200$ GeV.
The solid (open) triangles represent the coalescence model predictions
with FO1 (FO2).
The data point of protons from STAR measurement has been scaled by a
factor of $0.6$ to correct the weak decay effects~\cite{Ada04}.}
\label{YieldsFig}
\end{figure}

Figure~\ref{YieldsFig} shows the differential invariant yields
($d^2N/(2\pi p_Tdp_Tdy)$) of (anti-)nuclei evaluated at the transverse
momentum $p_T/|B|=0.875$ GeV/c as a function of baryon number $B$.
One can see that the coalescence model with FO1 reproduces the
measured differential invariant yields of p ($\overline{\text{p}}$),
d ($\overline{\text{d}}$) and $^3$He ($^3\overline{\text{He}}$)
very well but significantly underestimates the measured value of
$^4$He ($^4\overline{\text{He}}$), as already observed in
Fig.~\ref{Spect}. The dashed lines in Fig.~\ref{YieldsFig}
are obtained by fitting the differential invariant yields of
p ($\overline{\text{p}}$), d ($\overline{\text{d}}$) and
$^3$He ($^3\overline{\text{He}}$) by using an exponential
function $e^{-r|B|}$.
It is seen that
the differential invariant yields of p ($\overline{\text{p}}$),
d ($\overline{\text{d}}$) and $^3$He ($^3\overline{\text{He}}$)
follow the exponential function very well, depicting the same
exponential reduction rate of the differential invariant yields
with the increased atomic mass number for p ($\overline{\text{p}}$),
d ($\overline{\text{d}}$) and $^3$He ($^3\overline{\text{He}}$).
For $^4$He ($^4\overline{\text{He}}$),
however, the measured differential invariant yields significantly
deviate from the exponential function although the coalescence
model prediction with FO1 still follows the same exponential reduction rate.
For $^5$Li ($^5\overline{\text{Li}}$) and $^6$Li ($^6\overline{\text{Li}}$),
the predicted differential invariant yields with FO1 deviate the
exponential reduction rate by an enhancement factor of about $2.25$ $(3.57)$ and $1.87$ $(3.31)$
whereas those with FO2 display a much stronger enhancement by a
factor of about $16.6$ $(26.3)$ and $23.3$ $(41.2)$, respectively,
indicating a very strong binding energy effect.

Table~\ref{Yields} lists the $p_T$-integrated yield in the midrapidity region
($-0.5\le y \le 0.5$), i.e., $dN/dy$ at $y=0$, of light (anti-)nuclei.
One can easily obtain the midrapidity yield ratios
d/p ($\overline{\text{d}}$/$\overline{\text{p}}$)$=4.65\times 10^{-3}$ ($3.47\times 10^{-3}$) and
$^3\text{He}$/d ($^3\overline{\text{He}}$/$\overline{\text{d}}$)$=1.99\times 10^{-3}$ ($1.50\times 10^{-3}$).
In particular, if we take the midrapidity yield of $^4\text{He}$
($^4\overline{\text{He}}$) as the value predicted by the coalescence
model with FO2, we find the midrapidity yield ratios of $^5\text{Li}$/$^4\text{He}$
($^5\overline{\text{Li}}$/$^4\overline{\text{He}}$) and
$^6\text{Li}$/$^4\text{He}$ ($^6\overline{\text{Li}}$/$^4\overline{\text{He}}$)
increase, respectively, from $1.32\times 10^{-3}$ ($0.99\times 10^{-3}$) and
$1.67\times 10^{-6}$ ($0.94\times 10^{-6}$) with FO1 to $11.7\times 10^{-3}$
($8.74\times 10^{-3}$) and $26.4\times 10^{-6}$ ($14.9\times 10^{-6}$) with FO2.
These results indicate
that the binding energy effects can enhance the midrapidity yields of $^5$Li
($^5\overline{\text{Li}}$) and $^6$Li ($^6\overline{\text{Li}}$) by
a factor of about $8.77$ ($8.82$) and $15.9$ ($15.8$), respectively.

\begin{table}
\caption{$p_T$-integrated yield in the mid-rapidity ($-0.5\le y \le 0.5$) of light (anti-)nuclei.}
\begin{tabular}{c|c|c|c|c|c|c}
\hline \hline
           & p & d & $^3$He & $^4$He & $^5$Li & $^6$Li   \\
         \hline
        FO1 & 16.1 &7.49E-02 & 1.49E-04 & 1.54E-07 & 1.22E-09 & 1.53E-12         \\
         \hline
        FO2 & 16.1 & - &  -  & 9.18E-07 & 1.07E-8 & 2.43E-11         \\
         \hline
           & $\overline{\text{p}}$  & $\overline{\text{d}}$ & $^3\overline{\text{He}}$ & $^4\overline{\text{He}}$ & $^5\overline{\text{Li}}$ & $^6\overline{\text{Li}}$   \\
         \hline
        FO1 & 12.1 &4.21E-02 & 6.29E-05 & 4.88E-08 & 2.88E-10 & 2.73E-13         \\
         \hline
        FO2 & 12.1 & - &  - & 2.91E-07 & 2.54E-09 & 4.32E-12         \\
\hline \hline
\end{tabular}
\label{Yields}
\end{table}

\section{conclusion}

Based on the covariant coalescence model with a blast-wave-like
analytical parametrization for the (anti-)nucleon phase-space freezeout
configuration, we have extracted (anti-)nucleon freezeout information
in central Au+Au collisions at $\sqrt{s_{NN}} = 200$ GeV by fitting
the measured spectra of protons, deuterons and \text{$^3\text{He}$}.
We have found that the covariant coalescence model with the obtained
(anti-)nucleon phase-space freezeout configuration significantly
underestimates the measured yield of $^4\text{He}$ ($^4\overline{\text{He}}$).
We have shown
the predicted $^4$He yield can be enhanced to the
measured value by using a nucleon freezeout configuration in
which the nucleons are assumed to freeze out earlier than those
coalesced into deuterons and \text{$^3\text{He}$} to effectively
consider the large binding energy value of $^{4}$He. The similar
conclusion has been obtained for $^4\overline{\text{He}}$.

Assuming the similar binding energy effect also exists for the production
of heavier (anti-)\text{$^5\text{Li}$} and (anti-)\text{$^6\text{Li}$}
due to their comparable binding energy values with $^{4}$He, we have
predicted the spectra and yields of (anti-)\text{$^5\text{Li}$}
and (anti-)\text{$^6\text{Li}$} in central Au+Au collisions at
$\sqrt{s_{NN}} = 200$ GeV. Our results indicate that the binding
energy effect can significantly enhance the yields of
(anti-)\text{$^5\text{Li}$} and (anti-)\text{$^6\text{Li}$}.
In particular, the midrapidity yield ratios
$^5\text{Li}$/$^4\text{He}$ ($^5\overline{\text{Li}}$/$^4\overline{\text{He}}$) and
$^6\text{Li}$/$^4\text{He}$ ($^6\overline{\text{Li}}$/$^4\overline{\text{He}}$) increase,
respectively, from $1.32\times 10^{-3}$ ($0.99\times 10^{-3}$) and $1.67\times 10^{-6}$ ($0.94\times 10^{-6}$)
without binding energy effects to $11.7\times 10^{-3}$ ($8.74\times 10^{-3}$) and
$26.4\times 10^{-6}$ ($14.9\times 10^{-6}$) with binding energy effects.
Our results imply that
although the stable (anti-)$^6$Li nucleus is unlikely to be observed,
the unstable (anti-)$^5$Li nucleus could be produced in observable abundance
in ultrarelativistic heavy-ion collisions at RHIC where it may be identified
through the p-$^4\text{He}$ ($\overline{\text{p}}$-$^4\overline{\text{He}}$)
invariant mass spectrum.

Our present study suggests that the future experimental measurement on the
production of (anti-)\text{$^5\text{Li}$} in central Au+Au collisions
at $\sqrt{s_{NN}} = 200$ GeV would be extremely useful to test
the binding energy effect on the light (anti-)nuclei production, and thus
to understand the production mechanism of heavier
antimatter nuclei in ultrarelativistic heavy-ion collisions,
especially the observed enhancement for the yield of $^4$He
($^4\overline{\text{He}}$) compared to those of p ($\overline{\text{p}}$),
d ($\overline{\text{d}}$) and $^3$He ($^3\overline{\text{He}}$).
Any deviation of the measured (anti-)\text{$^5\text{Li}$} yield
in central Au+Au collisions at $\sqrt{s_{NN}} = 200$ GeV from the
coalescence model prediction with or without considering the binding
energy effect may indicate the exist of new excitation mechanism, e.g.,
the direct production of nuclei out of the highly correlated vacuum.

\begin{acknowledgments}
We are grateful to Che Ming Ko, Yu-Gang Ma and Zhang-Bu Xu for helpful discussions.
This work was supported in part by the Major State Basic
Research Development Program (973 Program) in China under Contract Nos.
2015CB856904 and 2013CB834405, the NNSF of China under Grant Nos. 11275125
and 11135011, the ``Shu Guang" project supported by Shanghai Municipal
Education Commission and Shanghai Education Development
Foundation, the Program for Professor of Special Appointment (Eastern Scholar)
at Shanghai Institutions of Higher Learning, and the Science and Technology
Commission of Shanghai Municipality (11DZ2260700).
\end{acknowledgments}


\begin{thebibliography}{99}

\bibitem{And33} C. D. Anderson, Phys. Rev. \textbf{43}, 491 (1933).

\bibitem{Dir28} P. A. M. Dirac, Proc. R. Soc. Lond. A  \textbf{117}, 610 (1928).

\bibitem{Cha55} O. Chamberlain, E. Segre, C. Wiegand, and T. Ypsilantis, Phys. Rev. \textbf{100}, 947 (1955).

\bibitem{Cor56} B. Cork, G. R. Lambertson, O. Piccioni, and W. A. Wenzel, Phys. Rev. \textbf{104}, 1193 (1956).

\bibitem{Mas65} T. Massam, T. Muller, B. Righini, M. Schneegans, and A. Zichichi, Nuovo Cim. \textbf{39}, 10 (1965).

\bibitem{Dor65} D. E. Dorfan, J. Eades, L. M. Lederman, W. Lee, and C. C. Ting, Phys. Rev. Lett. \textbf{14}, 1003 (1965).

\bibitem{Ant70} Y. M. Antipov et al., Yad. Fiz. \textbf{12}, 311 (1970).

\bibitem{Vis74} N. K. Vishnevsky et al., Yad. Fiz. \textbf{20}, 694 (1974).

\bibitem{Bau96} G. Baur et al., Phys. Lett. \textbf{B368}, 251 (1996).

\bibitem{ALPHA11} G. B. Andresen et al. (ALPHA Collaboration), Nature Phys. \textbf{7}, 558 (2011).

\bibitem{Abe10} B. I. Abelev et al. (The STAR Collaboration), Science \textbf{328}, 58 (2010).

\bibitem{Abe11} B. I. Abelev et al. (The STAR Collaboration), Nature \textbf{473}, 353 (2011).  

\bibitem{Sha11} N. Sharma, J. Phys. G \textbf{38}, 124189 (2011).

\bibitem{Ma12} Y. G. Ma, J. H. Chen, and L. Xue, Front. Phys. \textbf{7}, 637 (2012);
Y. G. Ma, J. Phys.: Conf. Series \textbf{420}, 012036 (2013); EPJ Web of Conf. \textbf{66}, 04020 (2014).

\bibitem{Ada15} J. Adam et al. (ALICE Collaboration), Nature Phys., (2015). DOI: 10.1038/NPHYS3432.

\bibitem{Ahl94} S. Ahlen et al., Nucl. Instrum. Methods \textbf{A350}, 351 (1994).

\bibitem{Fuk05} H. Fuke et al., Phys. Rev. Lett. \textbf{95}, 081101 (2005).

\bibitem{Don08} F. Donato, N. Fornengo, and D. Maurin, Phys. Rev. D \textbf{78}, 043506 (2008).

\bibitem{Abe12} K. Abe et al., Phys. Rev. Lett. \textbf{108}, 131301 (2012).

\bibitem{Ada15} L. Adamczyk et al. (The STAR Collaboration), Nature, in press, (2015) [arXiv:1507.07158].

\bibitem{Gre96} W. Greiner, Int. J. Mod. Phys. E \textbf{5}, 1 (1996); J. Phys.: Conf. Series \textbf{413}, 012002 (2013).

\bibitem{Xue12} L. Xue, Y. G. Ma, J. H. Chen, and S.Zhang, Phys. Rev. C \textbf{85}, 064912 (2012).

\bibitem{But61} S. T. Butler and C. A. Pearson, Phys. Rev. Lett. \textbf{7}, 69 (1961).
\bibitem{Sat81} H. Sato and K. Yazaki, Phys. Lett. \textbf{B98}, 153 (1981).
\bibitem{Cse86} L. P. Csernai and J. I. Kapusta, Phys. Rep. \textbf{131}, 223 (1986).

\bibitem{Cle91} J. Cleymans, K. Redlich, and E. Suhonen, Z. Phys. C \textbf{51}, 137 (1991).
\bibitem{Bec97} F. Becattini and U. W. Heinz, Z. Phys. C \textbf{76}, 269 (1997).
\bibitem{Bra07} P. Braun-Munzinger and J. Stachel, Nature \textbf{448}, 302 (2007).
\bibitem{And11} A. Andronic, P. Braun-Munzinger, J. Stachel, and H. Stoecker, Phys. Lett. \textbf{B697}, 203 (2011).
\bibitem{Cle11} J. Cleymans, S. Kabana, I. Kraus, H. Oeschler, K. Redlich, and N. Sharma, Phys. Rev. C \textbf{84}, 054916 (2011).


\bibitem{Zha10} S. Zhang, J. H. Chen, H. Crawford, D. Keane, Y. G. Ma, and Z. B. Xu, Phys. Lett. \textbf{B684}, 224 (2010).
\bibitem{ChenG12} G. Chen et al., Phys. Rev. C \textbf{86}, 054910 (2012).
\bibitem{ChenG13} G. Chen, H. Chen, J. Wu, D. S. Li, and M. J. Wang, Phys. Rev. C \textbf{88}, 034908 (2013).

\bibitem{Pop10} V. Topor Pop and S. Das Gupta, Phys. Rev. C \textbf{81}, 054911 (2010).
\bibitem{Ste12} J. Steinheimer, K. Gudima, A. Botvina, I. Mishustin, M. Bleicher, and H. Stoecker, Phys. Lett. \textbf{B714}, 85 (2012).
\bibitem{Cha14} S. Chatterjee and B. Mohanty, Phys. Rev. C \textbf{90}, 034908 (2014).
\bibitem{Cha15} S. Chatterjee, B. Mohanty, and R. Singh, Phys. Rev. C \textbf{92}, 024917 (2015).

\bibitem{Dov91} C. B. Dover, U. Heinz, and E. Schnedermann, Phys. Rev. C \textbf{44}, 1636 (1991).

\bibitem{Ret04} F. Reti\'{e}re and M. A. Lisa, Phys. Rev. C \textbf{70}, 044907 (2004).

\bibitem{Mat97} R. Mattiello, H. Sorge, H. Stoecker, and W. Greiner, Phys. Rev. C \textbf{55}, 1443 (1997).
\bibitem{ChenLW03} L. W. Chen, C. M. Ko, and B. A. Li, Phys. Rev. C \textbf{68}, 017601 (2003);
Nucl. Phys. \textbf{A729}, 809 (2003); Phys. Rev. C \textbf{69}, 054606 (2004).
\bibitem{ChenLW06} L. W. Chen and C. M. Ko, Phys. Rev. C \textbf{73}, 044903 (2006).
\bibitem{Oh09} Y. Oh, Z. W. Lin, and C. M. Ko, Phys. Rev. C \textbf{80}, 064902 (2009).

\bibitem{Coo74} F. Cooper and G. Frye, Phys. Rev. D \textbf{10}, 186 (1974).   

\bibitem{Sch99} R. Scheibl and U. Heinz, Phys. Rev. C \textbf{59}, 1585 (1999).

\bibitem{Bjo83} J. D. Bjorken, Phys. Rev. D \textbf{27}, 140(1983).

\bibitem{ChenLW0407} L. W. Chen, V. Greco, C. M. Ko, S. H. Lee, and W. Liu, Phys. Lett. \textbf{B601}, 34 (2004);
L. W. Chen, C. M. Ko, W. Liu, and M. Nielsen, Phys. Rev. C \textbf{76}, 014906 (2007).

\bibitem{Adl04} S. S. Adler et al.(The PHENIX Collaboration), Phys. Rev. C \textbf{69}, 034909 (2004).

\bibitem{Abe09} B. I. Abelev et al. (The STAR Collaboration), arXiv:0909.0566 [nucl-ex].



\bibitem{Rop09} G. Ropke, Phys. Rev. C \textbf{79}, 014002 (2009).

\bibitem{Tan85} I. Tanihata et al., Phys. Rev. Lett. \textbf{55}, 2676 (1985).

\bibitem{Wan12} M. Wang et al., Chin. Phys. C \textbf{36}, 1603 (2012).

\bibitem{Aud12} G. Audi et al., Chin. Phys. C \textbf{36}, 1157 (2012).


\bibitem{Ada04} J.Adams et al. (The STAR Collaboration), Phys. Rev. Lett. \textbf{92}, 112301 (2004).

\bibitem{Dan91} P. DanieIewicz and G. F. Bertsch, Nucl. Phys. \textbf{A533}, 712 (1991);
P. Danielewicz, Nucl. Phys. \textbf{A545}, 21c (1992); P. Danielewicz and Q. Pan, Phys. Rev. C \textbf{46}, 2002 (1992).


\end{thebibliography}
\end{document}